\documentclass[12pt,aps,preprint,a4paper]{revtex4}
\usepackage{amsmath}
\usepackage{graphicx}

\newcommand{\re}{\ref}
\newcommand{\be}{\begin{equation}}
\newcommand{\ee}{\end{equation}}

\newcommand{\la}{\label}
\newcommand{\ber}{\begin{eqnarray}}
\newcommand{\eer}{\end{eqnarray}}

\begin{document}
\title{A SMALL PARAMETER METHOD FOR FEW--BODY PROBLEMS\footnote{An extended version
of the talk \cite{1} at "Nucleus--2007", 24--29 June 2007, Voronezh, Russia}}

\author{V.D. Efros\footnote{E-mail efros@mbslab.kiae.ru}}

\affiliation{ RRC "Kurchatov Institute"}


\begin{abstract}
A procedure to solve few--body problems which is based on an expansion over a small 
parameter is developed. The parameter is the ratio of potential energy to kinetic energy 
in the subspace of states having not small hyperspherical quantum numbers, 
$K>K_0$.
Dynamic equations are reduced perturbatively to those in the finite subspace with 
$K\le K_0$. The contribution from the subspace with $K>K_0$ is taken into account in a 
closed form, i.e. without an expansion over basis functions.
\end{abstract}

\bigskip

\maketitle

\section{Introduction}
Below  an approach to solving few--body problems which is based on an 
expansion over a small parameter is developed. The parameter is the ratio of 
potential energy to kinetic energy for the states with hyperspherical numbers $K$ exceeding some limiting 
value $K_0$. Roughly speaking, the parameter is $K_0^{-2}$. The method is a
development of that of Ref. \cite{2}. An expansion over the parameter $K_0^{-2}$ has been given there
for solving large systems of linear equations that arise in bound--state problems in the framework of
the hyperspherical--hyperradial expansion.\footnote{In an early calculation \cite{3}
the wave function component with $K=2$ was obtained perturbatively from that with $K=0$. However,
the change of the component with $K=0$ itself due to coupling to $K=2$ has not been taken into
account. This missed quantity is in general of the same magnitude as that accounted for.} 
The method \cite{2} is efficient 
for this purpose \cite{4,5}. However, for A$>$3 it is the calculating of matrix elements entering those 
systems of equations that requires a massive computational effort. The difficulty stems from a swift rise 
of a number of hyperspherical states with the same $K$ as $K$ increases, or a number of particles
increases. Selection of hyperspherical states to reduce the effort, see \cite{2,6,7}, is 
efficient for A=3 and 4 bound--state problems only. Such a selection is not justified in reaction 
calculations, in particular. The problem is removed in the method below since no expansion over
basis states is employed here for $K>K_0$.

Recently a considerable progress in methods for solving few--body problems has been achieved.
However, those developments have limitations, and the latter are removed in the present method.
In particular, the well--known Green Function Monte Carlo method to be mentioned in this connection 
is the method to calculate a bound state of a system, and it is not suit to calculate
reactions. (Although the simplest scattering problems may be considered in it frames.) Besides,
this method is not convenient in the respect that it provides separate observables, such as an 
energy or a size, as a result of a calculation but it does not provide the wave function of a bound
state that could be employed in subsequent calculations. Unlike this method, the method below is
suitable for calculating reactions of a general type. And when in its frames one needs to use a 
bound state wave function one need not recalculate it completely each time.

Recently a way was found to extend the Faddeev--Yakubovsky A=4 calculations over the energy range above 
the four--body breakup threshold \cite{8}. However,  Yakubovsky type calculations require too much 
numerical effort even in the A=4 case. Amount of calculations is considerably less in the scheme 
below.

At solving few--body problems with expansion methods convergence of expansions for calculated 
quantities was accelerated with the help of the effective interaction approaches.
Such  approaches were developed in the framework of the oscillator expansion  \cite{9} and
the hyperspherical expansion \cite{10}. In their framework a true Hamiltonian is replaced with
some effective Hamiltonian acting in a subspace of only low excitations. When, formally, the
latter subspace is enlarged up to coincidence with the total space an effective Hamiltonian
turns to a true one. An effective Hamiltonian is constructed from a requirement that its ingredients, 
as defined in a subspace of low excitations, reproduce some properties of the corresponding
ingredients of a true Hamiltonian in the total space. It has been shown \cite{9,10} that this, indeed, 
leads to an improvement of convergence of observables considered.

Higher excitations are disregarded in such type calculations. It is clear, however, that correlation
effects related to higher excitations cannot be reproduced by any state vector lying in an allowed 
subspace of only low excitations. For example, let us consider the mean value, 
$\langle\Psi_0|H|\Psi_0\rangle$, of such an "observable" as a true Hamiltonian. 
It follows from the variational principle that an approximate state $\Psi_0$ supplied with such 
a method provides poorer approximation to the true  $\langle\Psi_0|H|\Psi_0\rangle$ value than 
$\Psi_0$ obtained by the simple diagonalization of a Hamiltonian in the same subspace of low 
excitations. And even the value of $\langle\Psi_0|H|\Psi_0\rangle$ obtained with the latter $\Psi_0$
is a very poor approximation for realistic  Hamiltonians. On the contrary, the method 
given below provides an approximate state vector  that is apparently close to a true state vector both 
as to its low excitation component and its high excitation component.

And speaking of reaction calculations in the framework of Eq. (\re{38}) below, 
$(H-\sigma){\tilde\Psi}=q$, one should in addition take into account that a rate of convergence is
determined not only by properties of the  Hamiltonian $H$ but also by those of the source--term $q$.
But these properties are apparently ignored at constructing effective  Hamiltonians. Therefore one
cannot expect fast convergence in all the cases, especially for source--terms $q$ corresponding to
strong--interaction induced reactions. On the contrary, the method described below provides
state vectors genuinely close to the true ones both for bound state problems and any reaction problems.

In the next section the bound state case is considered. In Sec. 3 modifications to treat reactions
are listed and a numerical estimate of the rate of convergence of the method is done. Some comments on 
computational aspects contain in Sec. 4.

\section{Bound states}

We consider the eigenvalue problem
\be
(H-E_\lambda)\Psi_\lambda =0,\la{1}
\ee
where $H=T+V$ is an A--body Hamiltonian. We 
split the whole space of states into the subspaces with $K\le K_0$ and $K>K_0$ and we denote 
$\Psi_\lambda^l$ and $\Psi_\lambda^h$ the components of the solution $\Psi_\lambda$ that lie, 
respectively, in these subspaces. At a proper choice of $K_0$ kinetic energy $T$ of a state belonging
to the second of the subspaces is much larger than its potential energy. Indeed,
\[
T=T_\rho+\frac{\hbar^2}{2m}\frac{{\hat K}^2}{\rho^2}, \]
where ${\hat K}^2$ is the hyperangular momentum operator acting on a hypersphere, $\rho$ is the
hyperradius, and $T_\rho$ is the hyperradial energy operator. The eigenvalues of the ${\hat K}^2$
operator are $K(K+n-2)$ where $n=3$A--3 is the dimension of a problem. Thus $\langle T\rangle$ is
large for states having large $K$ and not too large space extension. We choose $K_0$ in a way that
for $K>K_0$ one has, in a rough sense,
\be \left|\frac{\hbar^2}{2m}\frac{K^2}{\rho^2}\left[\Psi_\lambda\right]_K\right|\gg 
\left|\left[(V+T_\rho-E_\lambda)\Psi_\lambda\right]_K\right|.\la{2}\ee
Here $[\ldots]_K$ denotes a component of a state with a given $K$. Eq. (\re{2}) is to be fulfilled 
for all configurations that contribute significantly to a solution. The corresponding $\rho$ values 
range within the configuration space extension of a solution. These  $\rho$ values are such that
$\rho^2$ is less than, or about, $A\langle r^2\rangle$, where $r$ is the single--particle
size of a system.

At these conditions one may express the component $\Psi_\lambda^h$ in terms of $\Psi_\lambda^l$
perturbatively and obtain equations for the latter component alone. Let us define projectors onto
the $K\le K_0$ and $K>K_0$ subspaces as $P_{K_0}$ and $Q_{K_0}$, respectively. Let us express formally
$\Psi_\lambda^h$ in terms of $\Psi_\lambda^l$:
\be
\Psi_\lambda^h=-\Gamma_{K_0}(E_\lambda)V\Psi_\lambda^l,\la{3}
\ee
where 
\be
 \Gamma_{K_0}(E)=\left[Q_{K_0}(H-E)Q_{K_0}\right]^{-1}\la{4}
\ee
is the Green function defined in the second subspace. It is taken into account in (\re{3}) that
kinetic energy is diagonal with respect to $K$. It is convenient to define $\Gamma_{K_0}$ as acting 
in the whole space and to rewrite it in the form
 \be
 Q_{K_0}\left[Q_{K_0}(H-E)Q_{K_0}\right]^{-1}Q_{K_0}.\la{5}
\ee
Substituting Eq. (\re{3}) into the relationship
\be
P_{K_0}\left[(H-E_\lambda)\Psi_\lambda^l+V\Psi_\lambda^h\right]=0\la{6}
\ee
one gets the equation for $\Psi_\lambda^l$ alone,
\be
P_{K_0}(H-E_\lambda)\Psi_\lambda^l=P_{K_0}V\Gamma_{K_0}(E_\lambda)V\Psi_\lambda^l.\la{7}
\ee
The quantity $P_{K_0}V\Gamma_{K_0}(E_\lambda)VP_{K_0}$ represents the exact effective 
interaction arising due to coupling of the complementary $K>K_0$
subspace to the $K\le K_0$ subspace. 

We shall solve Eq. (\re{7}) perturbatively. We write in (\re{5}) $H-E=L+U$ and we use an expansion 
\be
\Gamma_{K_0}(E)=G_{K_0}-G_{K_0}U(E)G_{K_0}+G_{K_0}U(E)G_{K_0}U(E)G_{K_0}-\ldots,\la{8}
\ee 
where $G_{K_0}=Q_{K_0}(Q_{K_0}LQ_{K_0})^{-1}Q_{K_0}$. With the choices  of $L$ below it 
has no non--zero 
matrix elements between the subspaces with  $K\le K_0$ and $K>K_0$, i.e. $[Q_{K_0},L]=0$. Then
\[G_{K_0}=Q_{K_0}L^{-1}=L^{-1}Q_{K_0}.\]

For performing calculations in the coordinate representation we choose $G_{K_0}$ as follows,
\be
G_{K_0}=\left[\frac{\hbar^2}{2m}\frac{{\hat K}^2}{\rho^2}+W(\rho)\right]^{-1}Q_{K_0}.\la{9}
\ee
It  is convenient to represent (\re{9}) as a sum of contributions from various $K$ values,
\be
G_{K_0}=\sum_{K>K_0}g_K.\la{10}
\ee
Then
\be
\langle\xi|g_K|\xi'\rangle=\left[\frac{\hbar^2}{2m}\frac{K(K+n-2)}{\rho^2}+W(\rho)\right]^{-1}
\frac{\delta(\rho-\rho')}{\rho^{n-1}}\sum_\nu Y_{K\nu}^*({\hat \xi})Y_{K\nu}({\hat \xi}').
\la{11}\ee
Here $\xi$ and $\xi'$ are $n$--dimensional space vectors, $W(\rho)$ is a subsidiary interaction, 
${\hat \xi}=\xi/\rho$, ${\hat \xi}'=\xi'/\rho$, and $Y_{K\nu}$ form a complete set of orthonormalized
hyperspherical harmonics having the same $K$. The hyperangular factor entering here may be 
represented with the simple expression (e.g. \cite{11})
\be
\sum_\nu Y_{K\nu}^*({\hat \xi})Y_{K\nu}({\hat \xi}')=\frac{K+\gamma}{2\cdot\pi^{n/2}}
\Gamma(\gamma)C_K^\gamma({\hat \xi}\cdot{\hat \xi}'),\la{12}
\ee
where $C_K^\gamma(x)$ is the Gegenbauer polynomial.

The choice (\re{11}) of $g_K$ is done to facilitate Monte--Carlo calculations of matrix elements.
At this choice one has in (\re{8})
\be
\langle\xi|U(E)|\xi'\rangle=\langle\xi|V|\xi'\rangle+\delta(\xi-\xi')\left[T_\rho-E-W(\rho)\right],
\la{13}\ee
\be
T_\rho=-\frac{\hbar^2}{2m}\left(\frac{d^2}{d\rho^2}+\frac{n-1}{\rho}\frac{d}{d\rho}\right).\la{14}
\ee

To perform calculations in the momentum representation we suggest the expansion (\re{8}) with
a modified $G_{K_0}$, $G_{K_0}=[\Pi^2/(2m)-E_0+W(\Pi)]^{-1}Q_{K_0}$,
\ber
\langle{\bar\pi}|G_{K_0}|{\bar\pi}'\rangle=\frac{\delta({\bar\pi}-{\bar\pi}')-\Pi^{-(n-1)}\delta(\Pi-\Pi')
\sum_{K\le K_0;\nu}Y_{K\nu}^*({\bar \pi})Y_{K\nu}({\bar \pi}')}
{\Pi^2/(2m)-E_0+W(\Pi)}\nonumber\\
\equiv
\frac{\delta({\bar\pi}-{\bar\pi}')-\Pi^{-(n-1)}\delta(\Pi-\Pi')(2\cdot\pi^{n/2})^{-1}
\sum_{K\le K_0}(K+\gamma)C_K^\gamma({\hat \pi}\cdot{\hat \pi}')}
{\Pi^2/(2m)-E_0+W(\Pi)}\nonumber\\
=\frac{\Pi^{-(n-1)}\delta(\Pi-\Pi')(2\cdot\pi^{n/2})^{-1}
\sum_{K> K_0}(K+\gamma)C_K^\gamma({\hat \pi}\cdot{\hat \pi}')}
{\Pi^2/(2m)-E_0+W(\Pi)}\la{15},\eer
\be
\langle{\bar\pi}|U(E)|{\bar\pi}'\rangle=\langle{\bar\pi}|V|{\bar\pi}'\rangle
 -\delta({\bar\pi}-{\bar\pi}')[E-E_0+W(\Pi)].\la{16}
 \ee
Here $\bar\pi$ and $\bar\pi'$ are $n$--dimensional momentum vectors, $\Pi=|{\bar\pi}|$,
${\hat \pi}={\bar\pi}/\Pi$, ${\hat \pi}'={\bar\pi}'/\Pi$, and $W(\Pi)$ is a subsidiary
interaction. The quantity $E_0$ is a fixed energy chosen to be close to $E_\lambda$ sought for.

Roughly speaking, the expansion goes over $K_0^{-2}$. As $K_0$ increases relative contributions to 
a solution from subsequent terms in the expansion (\re{8}) decrease. Taking $K_0$ sufficiently large
we retain only the lower terms in the expansion. 

The subsidiary interaction $W(\rho)\simeq{\bar V(\rho)}$ or $W(\Pi)\simeq{\bar V(\Pi)}$ is intended to 
accelerate convergence of observables of interest when $K_0$ increases. A better choice of  
subsidiary interactions would be such that they include spin--isospin operators. Let us suppose
that calculations are performed in the coordinate representation. For a conventional
NN interaction that includes static local central and tensor components $V_{loc}$ plus components that 
depend on angular and linear momentum a possible good choice is the following. 
Let us consider the $K=0$ component in the expansion of $V_{loc}$ over hyperspherical harmonics. 
This component is the result of averaging $V_{loc}$ over a hypersphere. It has 
the structure $F(\rho){\hat O}$ where ${\hat O}=\sum{\hat O}(ij)$ is an operator that depends
on spin--isospin variables. The operator ${\hat O}$ is symmetric with respect to particle
permutations. Therefore it may be represented as 
\[{\hat O}=\sum_f O_f\sum_\mu \left|\theta^{[f]}_\mu\rangle\langle\theta^{[f]}_\mu\right|,\]
where $f$ labels irreducible representations of the permutation group of A particles,
$\mu$ labels basis vectors belonging to a representation $[f]$, $\{\theta^{[f]}_\mu\}$ is
the corresponding orthonormalized set of basis functions, and $O_f$ is defined as follows,
\be \langle\theta^{[f]}_\mu|{\hat O}|\theta^{[f]'}_{\mu'}\rangle=
\delta_{ff'}\delta_{\mu\mu'}O_f.\la{extr}\ee 
We then choose $W$ as 
\[ 
W=F(\rho)\sum_f O_f\sum_\mu \left|\theta^{[f]}_\mu\rangle\langle\theta^{[f]}_\mu\right|.\]
At this choice, $V$ and $W$ cancel each other to a large degree in the difference $V-W$ entering $U$.
This allows  employing a smaller $K_0$ value. 
The Green function $G_{K_0}=\sum_K g_K$ becomes
\ber
\langle\xi|g_K|\xi'\rangle=\frac{\delta(\rho-\rho')}{\rho^{n-1}}\frac{K+\gamma}{2\cdot\pi^{n/2}}
\Gamma(\gamma)C_K^\gamma({\hat \xi}\cdot{\hat \xi}')
\nonumber\\
\times\sum_f\left[\frac{\hbar^2}{2m}\frac{K(K+n-2)}{\rho^2}+F(\rho)O_f\right]^{-1}
\sum_\mu\left|\theta^{[f]}_\mu\rangle\langle\theta^{[f]}_\mu\right|.
\la{extr1}
\eer
The quantities $O_f$ may also be varied around their values from (\re{extr}). 
To simplify the presentation we did not include a spin--isospin 
dependence in the formulas above. 
  
We set in (\re{7})
\be
\Psi_\lambda^l=\sum_n\Psi_\lambda^{l(n)},\qquad E_\lambda=\sum_nE_\lambda^{(n)},\la{17}
\ee
where $\Psi_\lambda^{l(n)}$ and $E_\lambda^{(n)}$ correspond to the $n$--th order in the expansion over 
$G_{K_0}U$ in (\re{8}). We then get from (\re{7}), (\re{8})
\be 
P_{K_0}(H-E_\lambda^{(0)})\Psi_\lambda^{l(0)}=0,\la{18}
\ee
\be
P_{K_0}(H-E_\lambda^{(0)})\Psi_\lambda^{l(1)}=E_\lambda^{(1)}\Psi_\lambda^{l(0)}+
P_{K_0}VG_{K_0}V\Psi_\lambda^{l(0)},\la{19}
\ee
\ber 
P_{K_0}(H-E_\lambda^{(0)})\Psi_\lambda^{l(2)}=E_\lambda^{(1)}\Psi_\lambda^{l(1)}+
E_\lambda^{(2)}\Psi_\lambda^{l(0)}-P_{K_0}VG_{K_0}U(E_\lambda^{(0)})G_{K_0}V\Psi_\lambda^{l(0)}
\nonumber\\
+P_{K_0}VG_{K_0}V\Psi_\lambda^{l(1)}.\la{20}
\eer

If $\Psi_\lambda^{l(1)}$ is a solution to Eq. (\re{19}) then 
$\Psi_\lambda^{l(1)}+c\Psi_\lambda^{l(0)}$ with an arbitrary $c$ is also a solution. The same holds 
true as to $\Psi_\lambda^{l(2)}$ in (\re{20}). To get a unique solution it is sufficient to impose 
the normalization condition 
\be
\langle\Psi_\lambda^{l}|\Psi_\lambda^{l}\rangle=\langle\Psi_\lambda^{l(0)}|\Psi_\lambda^{l(0)}\rangle.
\la{21}\ee
This gives in the first and second order, respectively, 
\be
\langle\Psi_\lambda^{l(1)}|\Psi_\lambda^{l(0)}\rangle+
\langle\Psi_\lambda^{l(0)}|\Psi_\lambda^{l(1)}\rangle=0,\la{22}
\ee
\be
\langle\Psi_\lambda^{l(2)}|\Psi_\lambda^{l(0)}\rangle+
\langle\Psi_\lambda^{l(0)}|\Psi_\lambda^{l(2)}\rangle+
\langle\Psi_\lambda^{l(1)}|\Psi_\lambda^{l(1)}\rangle=0.\la{23}
\ee
Taking into account time reversal invariance of the operators entering (\re{19}), (\re{20}) it is
seen that the matrix elements in (\re{22}) and (\re{23}) are real. Therefore (\re{22}) and (\re{23})
turn to
\be
\langle\Psi_\lambda^{l(0)}|\Psi_\lambda^{l(1)}\rangle=0,\la{24}
\ee
\be
\langle\Psi_\lambda^{l(0)}|\Psi_\lambda^{l(2)}\rangle+
(1/2)\langle\Psi_\lambda^{l(1)}|\Psi_\lambda^{l(1)}\rangle=0.\la{25}
\ee

Taking scalar products of Eq. (\re{19}) and Eq. (\re{20}) with $\Psi_\lambda^{l(0)}$ 
and making use of Eq. (\re{18}) we obtain,
respectively,
\be
E_\lambda^{(1)}=-\frac{\langle\Psi_\lambda^{l(0)}|VG_{K_0}V|\Psi_\lambda^{l(0)}\rangle}
{\langle\Psi_\lambda^{l(0)}|\Psi_\lambda^{l(0)}\rangle},\la{26}
\ee 
\be
E_\lambda^{(2)}=\frac{\langle\Psi_\lambda^{l(0)}|VG_{K_0}U(E_\lambda^{(0)})G_{K_0}V
\Psi_\lambda^{l(0)}\rangle-\langle\Psi_\lambda^{l(1)}|VG_{K_0}V|\Psi_\lambda^{l(1)}\rangle}
{\langle\Psi_\lambda^{l(0)}|\Psi_\lambda^{l(0)}\rangle}.\la{27}
\ee
To get (\re{27}) Eq. (\re{24}) was employed.

We seek for the component $\Psi_\lambda^{l}$ as an expansion over the hyperspherical basis. In the
coordinate representation,
\be
\Psi_\lambda^{l}(\rho,{\hat\xi},\sigma_{zi},\tau_{zi})=\sum_{K\le K_0;\nu}\chi_{K\nu}(\rho)F_{K\nu}
({\hat\xi},\sigma_{zi},\tau_{zi}).\la{28}
\ee
Here $\sigma_{zi}$ and $\tau_{zi}$ are particle spin--isospin variables, $F_{K\nu}$ are basis 
functions that we consider to be orthonormalized. They are combinations of basis hyperspherical 
harmonics and basis spin--isospin functions. It is implied here and below that all the summations 
over $K$ include
only $K$ values of a given parity. Let us write down similar expansions for $\Psi_\lambda^{l(n)}$,
\[
\Psi_\lambda^{l(n)}(\rho,{\hat\xi},\sigma_{zi},\tau_{zi})=\sum_{K\le K_0;\nu}\chi_{K\nu}^{(n)}
(\rho)F_{K\nu}
({\hat\xi},\sigma_{zi},\tau_{zi}),\]
so that $\chi_{K\nu}=\sum_n\chi_{K\nu}^{(n)}$. Eqs. (\re{18}), (\re{19}) turn into equations for the
expansion coefficients $\chi_{K\nu}^{(n)}$:
\be (T_K-E_\lambda^{(0)})\chi_{K\nu}^{(0)}+\sum_{K'\le K_0;\nu'}(K\nu|V|K'\nu')\chi_{K'\nu'}^{(0)}=0.
\la{29}\ee
\be (T_K-E_\lambda^{(0)})\chi_{K\nu}^{(1)}+\sum_{K'\le K_0;\nu'}(K\nu|V|K'\nu')\chi_{K'\nu'}^{(1)}
=(K\nu|VG_{K_0}V|\Psi_\lambda^{l(0)})+E_\lambda^{(1)}\chi_{K\nu}^{(0)}.\la{30}\ee
\ber (T_K-E_\lambda^{(0)})\chi_{K\nu}^{(2)}+\sum_{K'\le K_0;\nu'}(K\nu|V|K'\nu')\chi_{K'\nu'}^{(2)}
\nonumber\\
=(K\nu|VG_{K_0}V|\Psi_\lambda^{l(1)})-(K\nu|VG_{K_0}U(E_\lambda^{(0)})G_{K_0}V|\Psi_\lambda^{l(0)})
+E_\lambda^{(1)}\chi_{K\nu}^{(1)}+E_\lambda^{(2)}\chi_{K\nu}^{(0)}.\la{31}\eer
Here $T_K$ denotes the hyperradial operator of kinetic energy,
\be 
T_K=T_\rho+\frac{\hbar^2}{2m}\frac{K(K+n-2)}{\rho^2}.\la{32}
\ee
In the notation above $(K\nu|\ldots)\equiv(F_{K\nu}|\ldots)$ and 
$(K\nu|V|K'\nu')\equiv(F_{K\nu}|V|F_{K'\nu'})$. These quantities are defined in a obvious way.
We recall that the equations written down include $K$ values only within a finite range, $K\le K_0$.
The zero order equations (\re{29}) are the standard ones that arise when coupling to states with 
$K>K_0$ is disregarded. The higher order equations just take this coupling into account.

Eq. (\re{22} reads as
\be 
\int\rho^{n-1}d\rho\sum_{K'\le K_0;\nu'}\chi_{K\nu}^{(0)}(\rho)\chi_{K\nu}^{(1)}(\rho)=0.\la{33}
\ee
The condition (\re{33}) is to be added to Eqs. (\re{30}). Let us suppose that Eqs. (\re{29}) and
(\re{30}) are solved via an expansion of $\chi_{K\nu}^{(0)}$ and $\chi_{K\nu}^{(1)}$ over the same
hyperradial basis with the same number of basis functions retained. The linear equations arising in 
this case from Eqs. (\re{30}) are linearly dependent. In general, one should remove one of these 
equations and 
replace it with the linear equation to which Eq. (\re{33}) turns. Eq. (\re{23}) becomes
\be 
\int\rho^{n-1}d\rho\sum_{K'\le K_0;\nu'}\chi_{K\nu}^{(0)}(\rho)\chi_{K\nu}^{(2)}(\rho)
+(1/2)\left[\chi_{K\nu}^{(1)}(\rho)\right]^2=0.\la{34}
\ee
This should be used similar to Eq. (\re{33}). If instead of (\re{28}) a hyperspherical expansion 
is employed within a momentum representation calculation similar equations may be written down proceeding 
from (\re{18})--(\re{20}).

The complementary $K>K_0$ component $\Psi_\lambda^{h}$ of a state sought for may be written as 
\be
\Psi_\lambda^{h}=\sum_n\Psi_\lambda^{h(n)}\la{35}
\ee
where $\Psi_\lambda^{h(n)}$ signifies a contribution having the $n$--th order in $G_{K_0}U$,
and $\Psi_\lambda^{h(0)}=0$. Then one has
\be
\Psi_\lambda^{h(1)}=-G_{K_0}V\Psi_\lambda^{l(0)},\la{36}
\ee
\be
\Psi_\lambda^{h(2)}=-G_{K_0}V\Psi_\lambda^{l(1)}+G_{K_0}U(E_\lambda^{(0)})
G_{K_0}V\Psi_\lambda^{l(0)}.\la{37}
\ee
The component $\Psi_\lambda^{l}$ has been obtained above in the form of a hyperspherical expansion.
Therefore one may store it and use in various applications. The complementary component 
$\Psi_\lambda^{h}$ then may be reconstructed as a simple quadrature (\re{36}), (\re{37}). 

If, for example, $\Psi_\lambda$ is calculated up to the $n>1$ corrections, i.e. 
$\Psi_\lambda^{appr}=\Psi_\lambda^{l(0)}+\Psi_\lambda^{l(1)}+\Psi_\lambda^{h(1)}$, then 
the average energy 
${\bar E}_\lambda=\langle\Psi_\lambda^{appr}|H|\Psi_\lambda^{appr}\rangle/
\langle\Psi_\lambda^{appr}|\Psi_\lambda^{appr}\rangle$ differs from the exact $E_\lambda$ value
 in terms  only of the third order and higher in the expansion over $G_{K_0}$. In particular,
 the second order energy (\re{27}) is correctly reproduced with ${\bar E}_\lambda$. 
 Indeed, according to the variational principle the difference between ${\bar E}_\lambda$
 and the exact $E_\lambda$ 
 value   includes the term 
 $\langle \delta\Psi_\lambda|H|\delta\Psi_\lambda\rangle$
 and powers of the term $\langle \delta\Psi_\lambda|\delta\Psi_\lambda\rangle$.
  Here $\delta\Psi_\lambda=
 \Psi_\lambda^{exact}-\Psi_\lambda^{appr}$. We have $\delta\Psi_\lambda\sim (G_{K_0})^2$ while presence
of $H$ in the above matrix element changes the net power in $G_{K_0}$ from $(G_{K_0})^4$ to
$(G_{K_0})^3$.

Basing on Table 4 in Ref. \cite{4} one infers the following. When only the above considered $n=1$
correction
is retained the choice $K_0=14$ ensures the correct binding 
energy at the accuracy level better than 0.1 MeV in the A=4 bound state problem
with a realistic NN interaction that includes a strong 
core.
The net number of HH with $K\le 14$ entering the problem does not exceeds several hundreds which is 
acceptable.\footnote{One might think that binding energy is better reproduced with the present
method than other variables since the $n=1$ correction provides an accuracy up to the second, and
not first, order in $G_{K_0}$. But, on the other hand, one should realize that the binding energy
considered is a small difference of two large quantities, potential and kinetic energies, which
deteriorates the accuracy.}  

\section{Reactions}

1. We consider a dynamic equation of the form
\be (H-\sigma){\tilde\Psi}=q.\la{38}\ee
Here $\sigma$ is a subsidiary complex energy, and $q$ is a given state. Reaction amplitudes may be 
obtained from ${\tilde\Psi}(\sigma)$ in a simple way, see e.g. the review \cite{12}. The approach
extensively applied to perturbation induced reactions and proved to be very efficient. 
Any strong--interaction induced reactions can also be treated with this approach. 

The solution $\tilde\Psi$ is localized. Therefore the procedure quite similar to that described
above is applicable also here. One represents $\tilde\Psi$ as $\tilde\Psi^l+\tilde\Psi^h$
and obtains these components as sums of successive approximations, 
${\tilde\Psi}^l=\sum_n{\tilde\Psi}^{l(n)}$,  ${\tilde\Psi}^h=\sum_n{\tilde\Psi}^{h(n)}$, 
where the meaning of 
notation is the same as above. One has
\be 
P_{K_0}(H-\sigma){\tilde\Psi}^{l(0)}=P_{K_0}q,\la{39}
\ee
\be
P_{K_0}(H-\sigma){\tilde\Psi}^{l(1)}=
P_{K_0}(VG_{K_0}V{\tilde\Psi}^{l(0)}-VG_{K_0}q),\la{40}
\ee
\be 
P_{K_0}(H-\sigma){\tilde\Psi}^{l(2)}=
P_{K_0}\left[-VG_{K_0}U(\sigma)G_{K_0}V{\tilde\Psi}^{l(0)}
+VG_{K_0}V{\tilde\Psi}^{l(1)}+VG_{K_0}U(\sigma)G_{K_0}q\right].\la{41}
\ee
As above these equations may be rewritten as coupled equations for coefficients of the hyperspherical 
expansion in the coordinate or momentum representation. The complementary components 
${\tilde\Psi}^{h(n)}$
are obtained from ${\tilde\Psi}^{l(n)}$ as quadratures,
\be
{\tilde\Psi}^{h(1)}=-G_{K_0}V{\tilde\Psi}^{l(0)}+G_{K_0}q,\la{42}
\ee
\be
{\tilde\Psi}^{h(2)}=-G_{K_0}V{\tilde\Psi}^{l(1)}+G_{K_0}U(\sigma)
G_{K_0}V{\tilde\Psi}^{l(0)}-G_{K_0}U(\sigma)G_{K_0}q.\la{43}
\ee 
When, for example, it is sufficient to account for only the $n=1$
corrections in $\Phi(\sigma)=\langle{\tilde\Psi}(\sigma)|{\tilde\Psi}(\sigma)\rangle$
one need not calculate the ${\tilde\Psi}^{h(n)}$ components.  
 
To estimate roughly the required $K_0$ value we note that the large--distance decay of
${\tilde\Psi}$ in the configuration space is determined by the imaginary part of the wave vector 
$k=[(2m/\hbar^2)^2\sigma]^{1/2}$. Let us write $\sigma=\sigma_R+\sigma_I$ and denote 
$R=({\rm Im}k)^{-1}$. 
Let us suppose that a calculation is performed in the coordinate representation, and the
expressions (\re{10}), (\re{11}), (\re{12}) for $G_{K_0}$ and (\re{13}) for $U$ with $E=\sigma$
are used. Then similar to (\re{2}) one may estimate the required $K_0$ value from the condition
\be
\frac{\hbar^2}{2mR^2}\left(K_0+\frac{n-2}{2}\right)^2\gg|V+\langle T_\rho\rangle-\sigma_R|.\la{44}
\ee
A typical $\sigma_I$ value is 10 MeV, and a required range of $\sigma_R$ values is about the same as 
a range
of energies considered in a problem. When $\sigma_R$ is not too high Eq. (\re{44}) is fulfilled for 
acceptably low $K_0$ values. (We shall not discuss the point on a precise $V$ value to be put there.)

When, however, the quantity $\sigma_R$ is high the expansion (\re{8}) of the
Green function converges quickly only for large $K_0$ values. (The deceleration of convergence
is caused by both terms $T_\rho$ and $\sigma$ in $U$, while they may compensate each other only 
in part.) To speed up the convergence, one could remove the contribution  $T_\rho-\sigma$ from $U$
and to account for it in $g_K$. However, this would hamper Monte--Carlo integrations because of 
the rapidly changing hyperradial Bessel and Hankel functions that would enter $g_K$ in this case.
One may avoid these complications if one performs calculations in the momentum representation. In
this case one uses Eq. (\re{15}) for $G_{K_0}$ with $E_0=\sigma$, 
\ber
\langle{\bar\pi}|G_{K_0}|{\bar\pi}'\rangle=
\frac{\delta({\bar\pi}-{\bar\pi}')-\Pi^{-(n-1)}\delta(\Pi-\Pi')(2\cdot\pi^{n/2})^{-1}
\sum_{K\le K_0}(K+\gamma)C_K^\gamma({\hat \pi}\cdot{\hat \pi}')}
{\Pi^2/(2m)-\sigma+W(\Pi)}\nonumber\\
=\frac{\Pi^{-(n-1)}\delta(\Pi-\Pi')(2\cdot\pi^{n/2})^{-1}
\sum_{K> K_0}(K+\gamma)C_K^\gamma({\hat \pi}\cdot{\hat \pi}')}
{\Pi^2/(2m)-\sigma+W(\Pi)}\la{45}.\eer
Correspondingly, in Eq. (\re{16}) for $U$ one replaces $E-E_0$ with zero,
\be
\langle{\bar\pi}|U(E)|{\bar\pi}'\rangle=\langle{\bar\pi}|V|{\bar\pi}'\rangle
 -\delta({\bar\pi}-{\bar\pi}')W(\Pi).\la{46}
\ee
In this case the condition
\[
\frac{\hbar^2}{2mR^2}\left(K_0+\frac{n-2}{2}\right)^2\gg V\]
is sufficient for quick convergence of the expansion (\re{8}) for the Green function.
Considering the role of subsequent terms in (\re{8}) in this case one should take into
account that if a coordinate representation wave function $f(\xi)$ is localized within a
hyperradius $\rho$ then the momentum representation quantities $(Y_{K\nu}({\hat \pi})|f(\pi))$
are very small at the $\Pi$ values such that $\Pi\rho\ll K+(n-2)/2$. (Irrespective to the 
mentioned condition, the condition $\sigma_R>>V$ also leads to quick convergence of the 
expansion (\re{8}).) While Eqs. (\re{45}), (\re{46}) are required for performing calculations 
that involve high $\sigma_R$ values, these equations, of course, may be employed at low $\sigma_R$
values as well.

Let us perform a rough estimate of efficiency of the latter version of the approach. Let 
us consider the A=4 case and adopt the $K_0$ value equal 14. Let us estimate the relative role 
of the correction ${\tilde\Psi}^{(2)}(\sigma)$ with respect to ${\tilde\Psi}^{(1)}(\sigma)$. 
For this purpose 
let us compare the contributions of these corrections to the net transform (see \cite{12}) 
$\Phi(\sigma)$ that correspond to $K=16$. These contributions are 
$\Phi^{(1)}_{K=16}(\sigma)\equiv\langle{\tilde\Psi}^{(1)}_{K=16}|{\tilde\Psi}^{(1)}_{K=16}\rangle$
and
$\Phi^{(1)+(2)}_{K=16}(\sigma)\equiv\langle{\tilde\Psi}^{(1)}_{K=16}+{\tilde\Psi}^{(2)}_{K=16}|
{\tilde\Psi}^{(1)}_{K=16}+{\tilde\Psi}^{(2)}_{K=16}\rangle$. We take in (\re{45}) $W(\Pi)=0$ 
and perform 
the calculation in the coordinate representation. For estimate purposes we can assume that 
${\tilde\Psi}^{l}$ is given, and with its help ${\tilde\Psi}^{h(1)}$ and ${\tilde\Psi}^{h(2)}$
are subsequently calculated as 
\ber
{\tilde\Psi}^{h(1)}=-G_{K_0}V{\tilde\Psi}^{l}+G_{K_0}q,\nonumber\\
{\tilde\Psi}^{h(2)}=-G_{K_0}V{\tilde\Psi}^{h(1)}.\la{47}
\eer 
(These expressions are not the same as (\re{42}), (\re{43}) since  we consider   
${\tilde\Psi}^{l}$ to be known here.) Let $\chi_{K\nu}^{(1)}(\rho)$ and $\chi_{K\nu}^{(2)}(\rho)$
be the coefficients of expansions of ${\tilde\Psi}^{h(1)}$ and ${\tilde\Psi}^{h(2)}$ over hyperspherical
harmonics. Then Eq. (\re{47}) turns to
\[\chi_{K\nu}^{(2)}(\rho)=\int_0^\infty g_K^{(0)}(\rho,\rho')(V{\tilde\Psi}^{h(1)})_{K\nu}(\rho')
(\rho')^8d\rho',\]
where 
\[(V{\tilde\Psi}^{h(1)})_{K\nu}=(Y_{K\nu}|V{\tilde\Psi}^{h(1)}),\]
and the free motion Green function is
\[g_K^{(0)}(\rho,\rho')=(2m/\hbar^2)(i\pi/2)(\rho\rho')^{-\gamma}J_{K+\gamma}(\sigma\rho_<)
H^{(1)}_{K+\gamma}(\sigma\rho_>).\]
Here $\rho_<={\rm min}(\rho,\rho')$, $\rho_>={\rm max}(\rho,\rho')$, $\gamma=(n-2)/2$. One may also
write at the $K=K_0+2$ value
\be(V{\tilde\Psi}^{h(1)})_{K\nu}(\rho')=\sum_{K'>K_0\nu'}V_{K\nu,K'\nu'}(\rho')\chi_{K'\nu'}^{(1)}(\rho')
\simeq V_{K\nu,K\nu}(\rho')\chi_{K\nu}^{(1)}(\rho')\simeq V_{0,0}(\rho')\chi_{K\nu}^{(1)}(\rho').
\la{48}\ee
In what follows we omit the subscript $\nu$ and perform the estimate up to multiplicities in $\nu$
both in $\langle{\tilde\Psi}^{(1)}_{K=16}|{\tilde\Psi}^{(1)}_{K=16}\rangle$ and 
$\langle{\tilde\Psi}^{(1)}_{K=16}+{\tilde\Psi}^{(2)}_{K=16}|
{\tilde\Psi}^{(1)}_{K=16}+{\tilde\Psi}^{(2)}_{K=16}\rangle$. Thus we use
\[\chi_{K=16}^{(2)}(\rho)=\int_0^\infty g_{K=16}^{(0)}(\rho,\rho')
V_{0,0}(\rho')\chi_{K=16}^{(1)}(\rho')(\rho')^8d\rho'.\]
For $\chi_{K=16}^{(1)}(\rho)$ we adopt the model
\be
\chi_{K=16}^{(1)}(\rho)=\frac{1}{(\sigma_R-\sigma_0)+i\sigma_I}\frac{e^{ik\rho}}{(\rho+\rho_0)^4},
\la{49}\ee
where $\sigma_0=50$ MeV, and $(\hbar k)^2/(2m)\equiv\sigma=\sigma_R+i\sigma_I$.\footnote{The 
approximation done in the first inequality in (\re{48}) is applicable when clustering of a state
${\tilde\Psi}$ is not very pronounced within its extensions. If $R$ denotes a size of a cluster this 
means that $kR$ is not extremely small. This is true at the value $\sigma_I=10$ MeV we use and not 
extremely high $\sigma_R$.} The expression (\re{49}) ensures the correct asymptotics at large $\rho$ 
values. 
We set $\sigma_I=10$ MeV that is a good choice to invert the transform, and $\rho_0=2$ fm. We
take $V=\sum V(r_{ij})$ and we employ the Gaussian potential $V(r)=V_0\exp(-r^2/b^2)$ with the 
parameters $V_0=$67 MeV and $b=$1.5 fm that corresponds to a triplet potential reproducing the
scattering length and the effective range. 
\begin{figure}
\begin{center}
\includegraphics[scale=0.5]{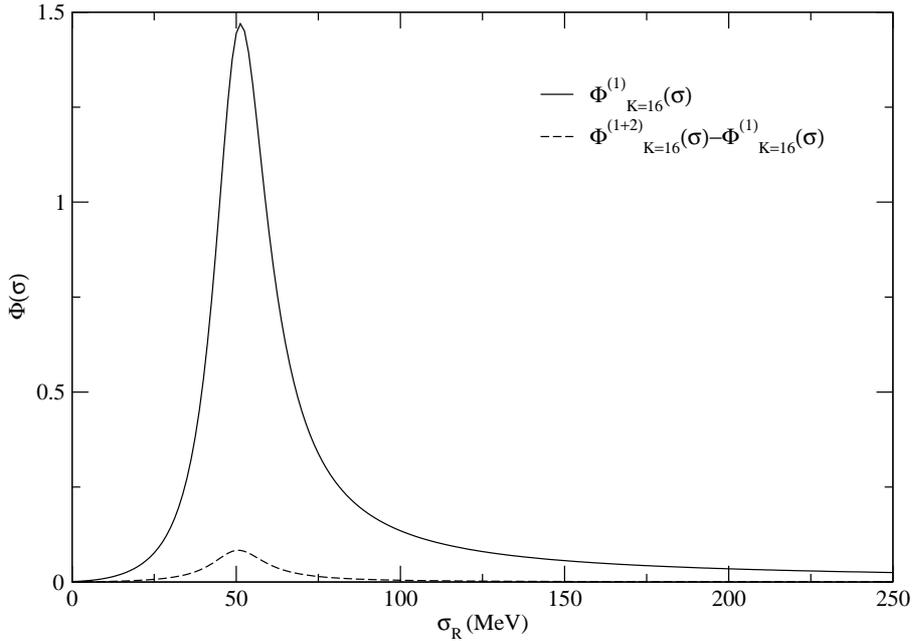}
\caption{The relative effect of the first and the second order corrections to the Lorentz transform
$\Phi(\sigma)=\langle{\tilde\Psi}|{\tilde\Psi}\rangle$. One has $\sigma=\sigma_R+i\sigma_I$, and
the $\sigma_I$ value is taken to be 10 MeV.}
\end{center}
\end{figure}
The values of $\Phi^{(1)}_{K=16}(\sigma)$ and 
$\Phi^{(1)+(2)}_{K=16}(\sigma)$ are shown in Fig. 1 as functions of $\sigma_R$.
It is seen that the second order correction to ${\tilde\Psi}$ is, indeed, of minor importance as 
compared to the first order correction.

2. Let us also consider reactions at low energy when only two--fragment channels are open. Suppose
that the dynamic equation $(H-E)\Psi_i=0$ for continuum spectrum states $\Psi_i$ is solved and $N$
channels are open. One may use the well--known ansatz
\be
\Psi_i=\phi_i^{(1)}+\sum_{j=1}^Nf_{ij}\phi_j^{(2)}+X,\la{50}
\ee
where $\phi_i^{(1)}$ and $\phi_j^{(2)}$ represent the "channel" states of two possible types, while $X$ 
is localized and is sought for as an expansion over hyperspherical harmonics. Let $K$ values up to
some $K_{max}$ are retained in the expansion, and $P_{K_{max}}$ is the projection operator onto the
subspace of those harmonics. The corresponding approximate equations include those to determine the 
expansion coefficients $\chi_{K\nu}(\rho)$ at reaction amplitudes $f_{ij}$ supposed to be "given". 
These equations may be written as
\be
P_{K_{max}}(H-E)X=P_{K_{max}}q,\la{51}
\ee
where 
\[q=-(H-E)\phi_i^{(1)}-\sum_{j=1}^Nf_{ij}(H-E)\phi_j^{(2)}.\]
To fix $f_{ij}$ one adds $N$ linear equations.\footnote{The corresponding exact equation $(H-E)X=q$
has a localized solution only when the reaction amplitudes $f_{ij}$ entering $q$ take their true values.
Therefore, unlike Eqs. (\re{51}), there is no need in $N$ additional equations here. On the contrary,
Eqs. (\re{51}) have a localized solution at any $f_{ij}$. When these $f_{ij}$ are different from the
true ones the components with high $K$ values reproduce a cluster structure of the solution so that 
there is no convergence in $K$.}

We note that Eqs. (\re{51}) may be solved perturbatively similar to Eq. (\re{38}). If $E<0$ is not too 
close to the three--fragment reaction threshold so that ${\bar \rho}\equiv\langle X|\rho|X\rangle$ is not too
large and the inequality (\re{44}) is fulfilled for moderate $K_0$ values then one can employ 
$K_0\ll K_{max}$ and retain only lower terms in the corresponding expansion over $G_{K_0}U$. (Of course,
this procedure is of limited use in the case of n--d scattering.) 

\section{Comments}

1. If the partial waves $\chi_{K\nu}(\rho)$ from (\re{28}) are sought for as an expansion over a set of
hyperradial functions the above dynamic equations turn to systems of linear equations. These systems may 
be of a large size. Then they can efficiently be solved with a version of the method of Ref. (\cite{2},
i.e. again using an expansion over another parameter of the $K_0^{-2}$ type. We note that when this 
method 
is applied subsequent iterations are identical  to each other in their form so that it is easy to perform a
required number of them. (Their number may typically be about ten or so to provide an accurate solution 
\cite{4}.) On the contrary, in the method described here an increase in a number of iterations  means
an increase in the dimension of the corresponding integrals. Therefore the present method is practical 
only when low order corrections are sufficient. This is the price for the elimination when $K>K_0$
of an expansion over basis states in the present 
method.

2. In the case of reaction calculations one passes from solutions of Eqs. (\re{38}) to reaction 
amplitudes 
as follows (see e.g. \cite{12}). The quantities of the 
$\Phi(\sigma)=\langle{\tilde\Psi}(\sigma)|{\tilde\Psi}(\sigma)\rangle$  type are formed. 
These quantities are 
integral transforms of response--like form factors that determine reaction amplitudes. So, to pass to 
reaction amplitudes these integral transforms are to be inverted. To perform a satisfactory inversion 
of the transform $\Phi(\sigma)$ one needs to use its values in rather many $\sigma$ points. But one need 
not 
solve 
Eqs. (\re{38}) for the corresponding many $\sigma$ values employed. A better approach is to solve these 
equations 
for rather a small number of $\sigma$ values and to obtain $\Phi(\sigma)$ for a larger set of $\sigma$
values via interpolation. The transforms $\Phi(\sigma)$ are smooth functions and this procedure is
safe and accurate.

3. Those matrix elements in the above equations which are related to contributions from $K>K_0$
are to be calculated with the Monte--Carlo method. It still should be verified numerically whether the
Monte--Carlo integration is efficient enough for this purpose. The existing experience testifies to that 
the Monte--Carlo integration is suitable at least in the case of matrix elements entering Eqs. 
(\re{29}) even when $K$ values are rather large (see e.g. \cite{13}).

The coordinate representation matrix elements above that correspond to the $n=1$ correction have the 
structure
\be
\sum_{K>K_0} \frac{K+\gamma}{2\cdot\pi^{n/2}}\Gamma(\gamma)\int \rho^{n-1}d\rho d{\hat \xi}d{\hat \xi}'
F_1(\rho{\hat \xi})\left[\frac{\hbar^2}{2m}\frac{K(K+n-2)}{\rho^2}+W(\rho)\right]^{-1}
C_K^\gamma({\hat \xi}\cdot{\hat \xi}')F_2(\rho{\hat \xi}').\la{52}
\ee  
Here, for example, $F_2(\rho{\hat \xi}')\equiv F_2(\xi')=\langle\xi'|V\Psi_\lambda^{l(0)}\rangle$.
The notation $d{\hat \xi}$ or $d{\hat \xi}'$ refers to hyperangular integrations. One deals with 
similar type integrals also in calculations that involve Eq. (\re{36}). Apart from a direct 
Monte--Carlo integration, in some cases it is expedient to take the argument of the Gegenbauer 
polynomial in (\re{52}) as a new variable, see Appendix, and to integrate over this variable with use
of the regular Gauss--Gegenbauer quadratures. While all other integrations are to be 
done with the Monte--Carlo
method. This can also be done in the case of a momentum representation calculation. Last lines
in Eq. (\re{15}) or (\re{45}) are to be used in this case. 

It is convenient to use permutational symmetry of states to simplify calculations of the $n=1$ 
correction. For example, when one retains only a two--body force, $V=\sum_{i<j}V(ij)$, one can
write
\ber
\langle\Psi_1|VgV|\Psi_2\rangle=\frac{A(A-1)}{2}\biggl[\langle\Psi_1|V(12)gV(12)|\Psi_2\rangle
+2(A-2)\langle\Psi_1|V(12)gV(13)|\Psi_2\rangle\nonumber\\ +
\left.\frac{(A-2)(A-3)}{2}\langle\Psi_1|V(12)gV(34)|\Psi_2\rangle\right].
\la{53}\eer  
When a three--body force is retained similar relationships could be written as well. Eq. (\re{52})
is written up to spin--isospin variables. When  Green functions $G_{K_0}$ or
$g$ from (\re{53})  are  spin--independent it is 
convenient to include
the intermediate spin--isospin factor $\sum_\mu|\theta_\mu\rangle\langle\theta_\mu|\equiv I$
in them, where $\{\theta_\mu\}$ is a complete set of spin--isospin states, c.f. (\re{extr1}). 

4. Let us comment on the $n=2$ correction. Suppose that a conventional NN interaction is employed 
that includes local central 
and tensor components plus components depending on orbital and linear momentum. 
Contributions from local components of such an interaction to the $n=2$ correction
have the following structure in the case of the coordinate representation calculation,
\ber
\sum_{K>K_0}\sum_{K'>K_0}\frac{(K+\gamma)(K'+\gamma)\Gamma^2(\gamma)}{(2\cdot\pi^{n/2})^2}\nonumber\\
\times\int \rho^{n-1}d\rho d{\hat \xi}_1d{\hat \xi}_2d{\hat \xi}_3 
\left[\frac{\hbar^2}{2m}\frac{K(K+n-2)}{\rho^2}+W(\rho)\right]^{-1}
\left[\frac{\hbar^2}{2m}\frac{K'(K'+n-2)}{\rho^2}+W(\rho)\right]^{-1}\nonumber\\
\times F_1(\rho{\hat \xi}_1)C_K^\gamma({\hat \xi}_1\cdot{\hat \xi}_2)V(\rho{\hat \xi}_2)
C_K^\gamma({\hat \xi}_2\cdot{\hat \xi}_3)F_2(\rho{\hat \xi}_3).\la{54} 
\eer
When  $n=2$ corrections are retained in a calculation sufficient accuracy is provided already with 
rather small $K_0$ values. Then contributions to (\re{54}) only from not large $K$ and $K'$ are 
significant which facilitates the Monte--Carlo integration. It may also be noted that contributions
of (\re{54}) type with the above mentioned non--local components of NN interaction include operators 
acting on the Gegenbauer polynomials. To disregard these contributions is a good approximation in
many cases.

5. As one could infer from Table 4 in \cite{4} a $K_0$ value required to ensure convergence
is considerably smaller in the case of NN interaction with a super soft core than that in the case 
of NN interaction with a strong core. Therefore one probably could reduce a required $K_0$
value also via transformation of dynamic equations to a form that involves a $t$ matrix
rather than an NN  potential. (In the A=3 case such equations are the Faddeev integral equations 
but at A$>$3 there is no need to pass to the Yakubovsky type equations for this purpose.)

\bigskip

\appendix*

\section{}

When one takes ${\hat \xi}\cdot{\hat \xi}'$ in (\re{52}) as a new integration variable one needs 
to define the whole set of integration variables in a way that the integrand remains smooth. This can 
be done e.g. as follows. Let us express ${\hat \xi}=\{{\hat \xi}_1,\ldots,{\hat \xi}_n\}$ in 
terms of another unit vector ${\hat \eta}$, 
\[{\hat \xi}_i=\sum_{j=1}^ng_{ij}{\hat \eta}_j,\]
where $g_{ij}$ is an orthogonal matrix such that its first column is $g_{i1}={\hat \xi}_i'$ and
$g_{ij}$ is arbitrary otherwise. One then has 
${\hat \xi}\cdot{\hat \xi}'=\sum_{i,j}g_{i1}g_{ij}{\hat \eta}_j={\hat \eta}_1$. Let us parametrize the 
components of ${\hat \eta}$ as follows,
\[{\hat \eta}_1=\cos\varphi,\qquad  {\hat \eta}_j={\hat v}_{j-1}\sin\varphi,\qquad j=2,\ldots,n,\]
where ${\hat v}_{i}$ are components of a unit vector ${\hat v}$ on a hypersphere in a $n-1$--dimensional
subspace. Taking into account that
\[ d{\hat \xi}=d{\hat \eta}\equiv(\sin\varphi)^{n-2}d{\hat v}d\varphi\]
one then may rewrite the integral (\re{52}) as
\ber
 \sum_{K>K_0} \frac{K+\gamma}{2\cdot\pi^{n/2}}\Gamma(\gamma)
 \int \rho^{n-1}d\rho d{\hat \xi}'d{\hat v}(\sin\varphi)^{2\gamma}d\varphi
F_1(\rho{\hat \xi})\nonumber\\
\left[\frac{\hbar^2}{2m}\frac{K(K+n-2)}{\rho^2}+W(\rho)\right]^{-1}
C_K^\gamma(\cos\varphi)F_2(\rho{\hat \xi}'),\nonumber
\eer 
where the components of the $n$--dimensional unit vector ${\hat \xi}$ entering $F_1$ are
parametrized as follows,
\[{\hat \xi}_i={\hat \xi}_i'\cos\varphi+
\left(\sum_{j=2}^ng_{ij}({\hat \xi}'){\hat v}_{j-1}\right)\sin\varphi.\]
The integrations over $d\rho$, $d{\hat \xi}'$, and $d{\hat v}$ may be performed with the Monte--Carlo
method while the remaining integration over $d\varphi$ may be done with the help of regular 
quadratures.

\end{document}